\pdfoutput=1

\documentclass[runningheads]{llncs}
\usepackage[T1]{fontenc}
\usepackage{tikz}
\usepackage{dsfont}

\usepackage{graphicx} 
\usepackage{hyperref} 
\usepackage[firstpage]{draftwatermark} 

\usepackage{algorithm}

\usepackage{orcidlink}

\usepackage{graphicx}
\usepackage{amsmath}
\usepackage{amssymb}
\usepackage{wrapfig}
\usepackage{mdframed}
\usepackage{hyperref}
\hypersetup{%
	colorlinks=true,           
	allcolors=brown!70!black,   
	pdfstartview=Fit,          
	breaklinks=true,
	pdfauthor={Raven, Tzu-Han, Borzoo, Bernd},
	pdftitle={{Syntax-Guided Automated Program Repair for Hyperproperties}}
}

\usepackage{cleveref}
\usepackage{color}

\usepackage{stmaryrd}
\usepackage{amsfonts}
\usepackage{ marvosym }

\usepackage{bussproofs}
\usepackage{booktabs}
\usepackage{colortbl}

\usepackage{xcolor}
\usepackage{relsize}
\usepackage{listings}

\usepackage{subcaption}
\usepackage{todonotes}
\usepackage{graphicx}

\usepackage{stmaryrd}

\usepackage{tcolorbox}
\tcbuselibrary{skins}

\newtcolorbox{encodingBox}[1]
{
	colback=white,
	boxrule=0.8pt,
	colbacktitle=white,
	coltitle=black,
	colframe=black,
	fonttitle=\bfseries,
	enhanced,
	attach boxed title to top right={yshift=-3mm, xshift=-4mm},
	boxed title style={boxrule=0.8pt},
	title={#1},
	arc = 0pt,
	top=5pt,
	left=4pt,
	before skip= 0pt,
	after skip=10pt
}

\newtcolorbox{codeBox}
{
	colback=white,
	boxrule=0.8pt,
	colbacktitle=white,
	coltitle=black,
	colframe=black,
	arc = 0pt,
	top=0pt,
	bottom=0pt,
	left=0pt,
	right=0pt,
	before skip= 0pt,
	after skip=0pt
}

\newcommand\xqed[1]{%
	\leavevmode\unskip\penalty9999 \hbox{}\nobreak\hfill
	\quad\hbox{#1}}
\newcommand\demo{\xqed{$\triangle$}}

\providecommand{\ltlN}{\operatorname{%
		\tikz[baseline]{
			\draw[line width=.12ex]
			(0,.6ex) circle (.8ex);
}}}{}

{}

{}

\DeclareMathOperator{\ltlW}{\mathcal{W}}

\newcommand{\edas}{\mathsf{edas}}

\newcommand{\traceVars}{\mathcal{V}}
\newcommand{\ldot}{\mathpunct{.}}
\newcommand{\traceset}{\mathcal{T}}
\newcommand{\symPathSet}{\mathcal{P}}

\newcommand{\intSet}{\mathbb{Z}}
\newcommand{\nat}{\mathbb{N}}
\newcommand{\bool}{\mathbb{B}}

\newcommand{\prog}{\mathbb{P}}
\newcommand{\progg}{\mathbb{Q}}
\newcommand{\proggg}{\mathbb{S}}

\definecolor{c1}{HTML}{602273}
\definecolor{c2}{HTML}{F39F5A}
\definecolor{c3}{HTML}{AE445A}
\definecolor{c4}{HTML}{6e4909}

\colorlet{statementColor}{c3}

\newcommand{\myobserve}{\text{\color{statementColor}\texttt{observe}}}
\newcommand{\myif}{\text{\color{statementColor}\texttt{if}}}
\newcommand{\mywhile}{\text{\color{statementColor}\texttt{while}}}
\newcommand{\myskip}{\text{\color{statementColor}\texttt{skip}}}

\DeclareMathOperator{\myassign}{{\color{statementColor}=}}
\DeclareMathOperator{\mysemi}{{\color{statementColor}\fatsemi}}

\newcommand{\sem}[1]{\llbracket #1 \rrbracket}

\newcommand{\traces}{\mathit{Traces}}
\newcommand{\states}{\mathit{Stores}}
\newcommand{\symstates}{\mathit{SymStores}}
\newcommand{\sympaths}{\mathit{SymPaths}}
\newcommand{\obs}{\mathit{obs}}

\newcommand{\quant}{\mathds{Q}}

\newcommand{\calA}{\mathcal{A}}

\newcommand{\calL}{\mathcal{L}}

\newcommand{\calF}{\mathcal{F}}
\newcommand{\calE}{\mathcal{E}}

\newcommand{\expInt}{\calE_\intSet}
\newcommand{\expBool}{\calE_\bool}

\newcommand{\symTo}{\xrightarrow{{}_\mathit{sym}} }

\newcommand{\tool}{\texttt{HyRep}}

\usepackage{scalefnt}

\newcommand\ScaleExists[1]{\vcenter{\hbox{\scalefont{#1}$\exists$}}}
\newcommand\ScaleForall[1]{\vcenter{\hbox{\scalefont{#1}$\forall$}}}

\DeclareMathOperator*\bigexists{%
	\vphantom\sum
	\mathchoice{\ScaleExists{1.7}}{\ScaleExists{1.4}}{\ScaleExists{1}}{\ScaleExists{0.75}}}

\DeclareMathOperator*\bigforall{%
	\vphantom\sum
	\mathchoice{\ScaleForall{1.7}}{\ScaleForall{1.4}}{\ScaleForall{1}}{\ScaleForall{0.75}}}

\colorlet{command-color}{black!70}
\definecolor{dkcyan}{rgb}{0.1, 0.3, 0.3}
\definecolor{dkgreen}{rgb}{0,0.3,0}
\definecolor{dkblue}{rgb}{0,0.3,1.0}
\colorlet{comment-color}{black!50}

\lstdefinelanguage{code-lang}{
	keywords={def, repeat, return, if, then, else,or},
	keywordstyle=[1]\color{command-color},
	morekeywords=[2]{faultLocalization,instrument, symbolicExecution,SyGuS},
	keywordstyle=[2]\color{dkgreen},
	morekeywords=[3]{iterativeRepair},
	keywordstyle=[3]\color{dkcyan},
	comment=[l][\color{comment-color}]{//},
	literate=%
	{=}{{{\color{command-color}=}}}1
	{|}{{{\color{command-color}|}}}1
	{:}{{{\color{command-color}:}}}1
	{:=}{{{\color{command-color}:=}}}1
	{@}{ }1
}

\lstdefinestyle{code-style}{
	escapeinside={(*}{*)},
	basicstyle=\ttfamily\fontsize{9}{11}\selectfont,
	columns=fullflexible,
	commentstyle=\sffamily\color{black!50!white},
	framexleftmargin=1em,
	framexrightmargin=1ex,
	keepspaces=true,
	mathescape,
	numbers=left,
	numberblanklines=false,
	numbersep=0.5em,
	numberstyle=\relscale{0.65}\color{gray}\ttfamily,
	showstringspaces=true,
	stepnumber=1,
	xleftmargin=1.2em,
}

\lstdefinelanguage{example-lang}{
	keywords={while,do,if, then,else, observe,or},
	keywordstyle=[1]{\color{c3}},
	morekeywords=[2]{string,int,bool},
	keywordstyle=[2]\color{black!60},
	morekeywords=[3]{phase,title,session,decision,print,password,attack,request,username,date,userPassword,info,LOG,credentials,reviewA,reviewB, reviewerAid,reviewerBid,order,notification,withdraw,balance,amount,ErrorLog,TransactionLog},
	keywordstyle=[3]\color{c1},
	morekeywords=[4]{366,1,0,true},
	keywordstyle=[4]\color{c4},
	morestring=[b]",
	stringstyle=\color{c4},
	comment=[l][\color{comment-color}]{//},
	literate=%
	{=}{{{\color{c3}=}}}1
	{!=}{{{\color{c3}!=}}}1
	{>}{{{\color{c3}>}}}1
	{<}{{{\color{c3}<}}}1
	{+}{{{\color{c3}+}}}1
	{-}{{{\color{c3}-}}}1
	{*}{{{\color{c3}*}}}1
	{@}{ }1
}

\lstdefinestyle{example-style}{
	escapeinside={(*}{*)},
	basicstyle=\ttfamily\fontsize{8}{10}\selectfont,
	columns=fullflexible,
	commentstyle=\sffamily\color{black!50!white},
	framexleftmargin=0em,
	framexrightmargin=0ex,
	keepspaces=true,
	mathescape,
	numbers=left,
	numberblanklines=false,
	numbersep=1.0em,
	numberstyle=\relscale{0.65}\color{gray}\ttfamily,
	showstringspaces=true,
	stepnumber=1,
	xleftmargin=1.2em
}

\lstdefinestyle{example-style-large}{
	escapeinside={(*}{*)},
	basicstyle=\ttfamily\fontsize{10}{12}\selectfont,
	columns=fullflexible,
	commentstyle=\sffamily\color{black!50!white},
	framexleftmargin=0em,
	framexrightmargin=0ex,
	keepspaces=true,
	mathescape,
	numbers=left,
	numberblanklines=false,
	numbersep=0.5em,
	numberstyle=\relscale{0.65}\color{gray}\ttfamily,
	showstringspaces=true,
	stepnumber=1
}

\lstnewenvironment{code}[1][]
{\small
	\lstset{
		style=code-style, 
		language=code-lang,
		#1
	}
}
{}

\lstnewenvironment{exampleCode}[1][]
{\small
	\lstset{
		style=example-style, 
		language=example-lang,
		#1
	}
}
{}

\newcommand{\ex}[1]{\lstinline[style=example-style-large,language=example-lang,]|#1|}
\newcommand{\exm}[1]{\text{\lstinline[style=example-style-large,language=example-lang,]|#1|}}

\begin{document}
\title{Syntax-Guided Automated Program Repair \\ for Hyperproperties}
\titlerunning{Syntax-Guided Automated Program Repair for Hyperproperties}
%

\author{Raven Beutner\inst{1}\scalebox{1.2}{\orcidlink{0000-0001-6234-5651}} \and
	Tzu-Han Hsu\inst{2}\scalebox{1.2}{\orcidlink{0000-0002-6277-2765}} \and
	Borzoo Bonakdarpour\inst{2}\scalebox{1.2}{\orcidlink{0000-0003-1800-5419}} \and 
	Bernd Finkbeiner\inst{1}\scalebox{1.2}{\orcidlink{0000-0002-4280-8441}}}
\authorrunning{R.~Beutner et al.}
%

\institute{CISPA Helmholtz Center for Information Security, \\
	Saarbrücken, Germany \\
		\email{\{raven.beutner,finkbeiner\}@cispa.de}
	\and
	Michigan State University,  East Lansing, MI, USA\\
	\email{\{tzuhan,borzoo\}@msu.edu}
}

\maketitle              

\SetWatermarkAngle{0}
\SetWatermarkText{\raisebox{11.5cm}{%
		\hspace{0.1cm}%
		\href{https://doi.org/10.5281/zenodo.10947975}{\includegraphics{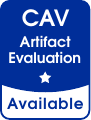}}%
		\hspace{9cm}%
		\includegraphics{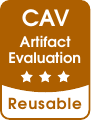}%
}}

\begin{abstract}
We study the problem of automatically repairing infinite-state software programs w.r.t.~temporal hyperproperties.
As a first step, we present a repair approach for the temporal logic HyperLTL based on symbolic execution, constraint generation, and 
syntax-guided synthesis of repair expression (SyGuS).
To improve the repair quality, we introduce the notation of a \emph{transparent} repair that aims to find a patch that is as 
close as possible to the original program.
As a practical realization, we develop an  \emph{iterative} repair approach.
Here, we search for a sequence of repairs that are closer and closer to the original program's behavior. 
We implement our method in a prototype and report on encouraging experimental results using off-the-shelf SyGuS solvers. 
\end{abstract}

\section{Introduction}
\label{sec:overview}

\emph{Hyperproperties} and \emph{program repair} are two popular topics within the formal 
methods community. 
Hyperproperties \cite{ClarksonS08} relate multiple executions of a system and occur, e.g., in information-flow 
control \cite{ZdancewicM03}, robustness 
\cite{ChaudhuriGL12}, and concurrent data structures~\cite{BonakdarpourSS18}. 
Traditionally, automated program repair (APR) \cite{GouesPR19,GazzolaMM19} attempts to repair 
the \emph{functional} behavior of a program.
In this paper, we, for the first time, tackle the challenging combination of APR and hyperproperties: given an 
(infinite-state) software program $\prog$ and a violated hyperproperty $\varphi$, repair $\prog$ 
such that $\varphi$ is satisfied.

As a motivating example, consider the data leak in the EDAS conference 
manager~\cite{AgrawalB16} (simplified in \Cref{fig:edas}).
The function \ex{display} is given the current phase of the review process (\ex{phase}), paper title (\ex{title}), session (\ex{session}), and 
acceptance decision (\ex{decision}), and computes a string (\ex{print}) that will be displayed to the author(s). 
As usual in a conference management system, the displayed string should not leak information other than the \ex{title}, unless the review process has been concluded. 
We can specify this {\em non-interference} policy as a hyperproperty in 
HyperLTL~\cite{ClarksonFKMRS14} as follows:
\begin{align}\label{eq:edas}
	\begin{split}
		\forall \pi_1\ldot \forall \pi_2\ldot \big( \exm{phase}_{\pi_1} &\!\neq \exm{"Done"} 
		\, \land \,  \,\exm{phase}_{\pi_2} \! \neq \exm{"Done"} \; \land\\
		&\exm{title}_{\pi_1} \!= 
		\exm{title}_{\pi_2} \big) \to \ltlN \big(\exm{print}_{\pi_1} \!=\exm{print}_{\pi_2}\big).
	\end{split}\tag{$\varphi_{\edas}$}
\end{align}
\begin{wrapfigure}{r}{0.49\textwidth}
	\vspace{-2mm}
\begin{exampleCode}
display(string phase, string title, 
@string session, string decision) {
@@@observe(*\label{line:observe1}*)
@@@decision = decision(*\label{line:repair-motivations}*)
@@@if (decision == "Accept") {
@@@@@@print = title + session
@@@} else {
@@@@@@print = title
@@@}
@@@observe(*\label{line:observe2}*)
}\end{exampleCode}
	\vspace{-2mm}
	\caption{Information leak in EDAS conference management 
		system. }\label{fig:edas}
	\vspace{-5mm}
\end{wrapfigure}
That is, for any \emph{two} execution traces $\pi_1, \pi_2$ of \ex{display} that, initially (i.e., at the first 
$\myobserve$ statement in line \ref{line:observe1}), have not reached the \ex{"Done"} phase (i.e., $\exm{phase}  \neq \exm{"Done"}$) and agree on the title, should, at the second $\myobserve$ in line \ref{line:observe2}, agree on the value of \ex{print}. 
It is straightforward to observe that function \ex{display} violates \ref{eq:edas}.
The code \emph{implicitly} leaks the acceptance decision by printing the session iff the paper is accepted.
A natural question to ask is whether it is possible to automatically {\em repair} the \ex{display} function such that \ref{eq:edas} is satisfied.

\paragraph{Constraint-Based Repair for Hyperproperties.}

As a first contribution, we propose a constraint-based APR approach for HyperLTL. 
Similar to existing constraint-based APR methods for functional properties \cite{NguyenQRC13}, we rely on fault localization to identify potential repair locations (e.g., line \ref{line:repair-motivations} of our example in \Cref{fig:edas}).
We then replace the repair locations with a fresh function symbol; use symbolic execution to explore symbolic paths of the program; and generate repair constraints on the inserted function symbols. 
We show that we can use the {\em syntax-guided synthesis} (SyGuS) framework \cite{AlurBJMRSSSTU13} to express (and solve) the repair constraints for HyperLTL properties with an \emph{arbitrary} quantifier prefix.

\paragraph{Many Solutions.}

The main challenge in APR for hyperproperties lies in the large number of possible repair patches; a problem that already exists when repairing against functional properties \cite{SmithBGB15} but is even more amplified when targeting hyperproperties. 
Different from functional specification, hyperproperties do not reason about the concrete functional (trace-level) behavior of a program, and rather express abstract relations between multiple computation traces. 
For example, information-flow policies such as observational determinism \cite{ZdancewicM03} can be checked and applied to arbitrary programs, regardless of their functional behavior.
In contrast to functional trace properties, we thus cannot partition the set of all program executions into ``correct'' executions (i.e., executions that already satisfy the trace property and should be preserved in the repair) and ``incorrect''  executions.
Instead, we need to alter the \emph{set} of all program executions such that the executions together satisfy the hyperproperty, leading to an even larger space of potential repairs. 
Moreover, within this large space, many repairs trivially satisfy the hyperproperty by severely changing the functional behavior of the program, which is usually not desirable.

In our concrete example, the \ref{eq:edas} property implicitly reasons about the (in)dependence 
between \ex{phase}, \ex{title}, and \ex{print} but does not impose \emph{how} the 
(in)dependence is realized functionally. 
If we apply our basic SyGuS-based repair approach, i.e., search for \emph{some} repair of line 
\ref{line:repair-motivations} that satisfies $\varphi_{\edas}$, it will immediately return 
a trivial repair patch: \exm{decision = "Reject"}.
This repair simply sets the \ex{decision} to some string not equal to \ex{"Accept"} (we use 
\ex{"Reject"} here for easier presentation).
While this certainly satisfies our information-flow requirement, it does not yield a desirable implementation of 
\ex{display} because the session is never displayed. 

\paragraph{Transparent Repair.}

To tackle this issue, we strengthen our repair constraints using the concept of \emph{transparency} (borrowed from the runtime enforcement literature~\cite{NgoMMP15}). 
Intuitively, we search for a repair that not only satisfies the hyperproperty but preserves as much functional behavior of the original program as possible.
We show that we can integrate this within our SyGuS-based repair constraints. 
In the extreme, \emph{full transparency} states that a repair is only allowed to deviate from the original program's behavior if absolutely necessary, i.e., only when the original behavior is part of a violation of the hyperproperty.

\paragraph{Iterative Repair.}

In the setting of hyperproperties, full transparency is often not particularly useful.
It strictly dictates what traces can be changed by a repair, 
potentially resulting in the absence of a repair (within a given search space).
In other instances (including the EDAS example), many paths (in the EDAS example, \emph{all} paths) take part in some violation of the hyperproperty, allowing the repair to intervene arbitrarily.
We introduce a more practical repair methodology that follows the same objective as (full) transparency (i.e., preserve as much original program behavior as possible).
Our method, which we call \emph{iterative repair}, approximates the global search for an optimal 
repair by a step-wise search for repairs of increasing quality.
Concretely, starting from some initial repair, we iteratively try to find repair patches that preserve \emph{more} original program behavior than our previous repair candidate. 
We show that we can effectively encode this into SyGuS constraints, and existing off-the-shelf SyGuS solvers can handle the resulting queries in many challenging instances. 
Notably, while some APR approaches (for functional properties) also try to find repairs that are close to the original program, they often do so heuristically. 
In contrast, our iterative repair constraints \emph{guarantee} that the repair candidates strictly improve in each iteration. 
See \Cref{sec:related-work} for more discussion. 

\begin{figure}[!t]
	\begin{subfigure}[b]{0.32\linewidth}
		\begin{codeBox}
			\vspace{-2mm}
\begin{exampleCode}[numbers=none, xleftmargin=0.4em]
if (phase == "Done"){
@@@decision = decision 
} else {
@@@decision = "Reject"
}\end{exampleCode}
\vspace{-2mm}
		\end{codeBox}
		\subcaption{}\label{fig:repair-1}
	\end{subfigure}\hfil
	\begin{subfigure}[b]{0.65\linewidth}
		\begin{codeBox}	
			\vspace{-2mm}
\begin{exampleCode}[numbers=none, , xleftmargin=0.3em]
if ((phase == "Done") or (decision != "Accept")){
@@@decision = decision 
} else {
@@@decision = "Reject"
}
\end{exampleCode}
\vspace{-2mm}
		\end{codeBox}
		\subcaption{}\label{fig:repair-2}
	\end{subfigure}
	\vspace{-1mm}
	\caption{Repair candidates discovered by our iterative repair.}\label{fig:repair-o}
\end{figure}

\paragraph{Iterative Repair in Action.}

Coming back to our initial EDAS example, we can use iterative repair to improve upon the na\"ive repair \exm{decision = "Reject"}. 
When using our iterative encoding, we find the improved repair solution in \Cref{fig:repair-1} that (probably) best mirrors the intuition 
of a programmer (cf.~\cite{PolikarpovaSYIH20}):
This repair patch only overwrites the decision in cases where the phase does not equal \ex{"Done"}.
In particular, note how our iterative repair finds the \emph{explicit} dependence of \ex{decision} on \ex{phase} (in the form of a conditional) even though this is only specified \emph{implicitly} in \ref{eq:edas}. 
In a third iteration, we can find an even closer repair, displayed in \Cref{fig:repair-2}:
This repair only changes the decision if the review process is not completed \emph{and} the decision equals \ex{"Accept"}.  

\paragraph{Implementation.}

We implement our repair approach in a prototype named \tool{} and evaluate \tool{} on a set of repair instances, including $k$-safety properties from the literature and challenging information-flow requirements.

\paragraph{Structure.} 
\Cref{sec:pre} presents basic preliminaries, including our simple programming language and the formal specification language for hyperproperties targeted by our repair.
\Cref{sec:simple-repair} introduces our basic SyGuS-based repair approach, and we discuss our transparent and 
iterative extensions in Sections~\ref{sec:trans-repair} and~\ref{sec:iterative}, respectively.
We present our experimental evaluation in~\Cref{sec:implementation} and discuss related work in~\Cref{sec:related-work}.

\section{Preliminaries}
\label{sec:pre}

Given a set $Y$, we write $Y^*$ for the set of finite sequences over $Y$, $Y^\omega$ for the set of infinite sequences, and $Y^\star := Y^* \cup Y^\omega$ for the set of finite and infinite sequences. 
For $t \in Y^\star$, we define $|t| \in \nat \cup \{\infty\}$ as the length of $t$.

\paragraph{Programs.}

Let $X$ be a fixed set of program variables.
We write $\expInt$ and $\expBool$ for the set of all arithmetic (integer-valued) and Boolean expressions over $X$, respectively. 
We consider a simple (integer-valued) programming language
\begin{align*}
	\prog, \progg := \myskip \mid x \myassign e \mid \myif(b, \prog, \progg) \mid \mywhile(b, \prog) \mid \prog \mysemi \progg  \mid \myobserve
\end{align*}
where $x \in X$, $e \in \expInt$, and $b \in \expBool$.
Most statements behave as expected. 
Notably, our language includes a dedicated $\myobserve$ statement, which we will use to express \emph{asynchronous} (hyper)properties \cite{GutsfeldMO21,BaumeisterCBFS21,BozzelliPS21}. 
Intuitively, each $\myobserve$ statement causes an observation in our temporal formula, and we skip over un-observed (intermediate) computation steps (see also \cite{BeutnerF22}). 

\paragraph{Semantics.}

Programs manipulate (integer-valued) stores $\sigma : X \to \intSet$, and we define $\states := \{ \sigma \mid \sigma : X \to \intSet  \}$ as the set of all stores.
Our (small-step) semantics operates on configurations $C = \langle \prog, \sigma\rangle$, where $\prog$ is a program and $\sigma \in \states$.
Reduction steps have the form $C \xrightarrow{\mu} C'$, where $\mu \in \states \cup \{\epsilon\}$.
Most program steps have the form $C \xrightarrow{\epsilon} C'$ and model a transition without observation. 
Every execution of an $\myobserve$ statement induces a transition $C \xrightarrow{\sigma} C'$, modeling a transition in which we observe the current store $\sigma$.
\Cref{fig:semantic-rules} depicts a selection of reduction rules.
For a program $\prog$ and store $\sigma$, there exists a unique \emph{maximal} execution $\langle \prog, \sigma \rangle \xrightarrow{\mu_1} \langle \prog_1, \sigma_1 \rangle \xrightarrow{\mu_2} \langle \prog_2, \sigma_2 \rangle \xrightarrow{\mu_3} \cdots$, where $\mu_1, \mu_2, \mu_3, \ldots \in \states \cup \{\epsilon\}$. 
Note that this execution can be finite or infinite. 
We define $\obs(\prog, \sigma) := \mu_1 \mu_2 \mu_3 \cdots \in \states^\star$ as the (finite or infinite) \emph{observation sequence} along this execution (obtained by removing all $\epsilon$s).
We write $\traces(\prog) := \{\obs(\prog, \sigma) \mid \sigma \in \states\} \subseteq \states^\star$ for the set of all traces generated by $\prog$.
We say a program $\prog$ is \emph{terminating}, if all its executions are finite. 

\begin{figure}[!t]
	
	\vspace{-3mm}
	\begin{minipage}{0.5\textwidth}
		\vspace{2.6mm}
		\small
				\def\defaultHypSeparation{\hskip .2in}
				\begin{prooftree}
					\AxiomC{}
					\UnaryInfC{$\langle x \myassign e, \sigma \rangle \xrightarrow{\epsilon} \langle \myskip, \sigma[x \mapsto \sem{e}_\sigma] \rangle$}
				\end{prooftree}
		\end{minipage}%
		\begin{minipage}{0.5\textwidth}
			\vspace{2.6mm}
			\small
					\def\defaultHypSeparation{\hskip .2in}
					\begin{prooftree}
						\AxiomC{}
						\UnaryInfC{$\langle \myobserve, \sigma \rangle \xrightarrow{\sigma} \langle \myskip,  \sigma \rangle$}
					\end{prooftree}
				\end{minipage}

		\vspace{3mm}
		
		\begin{minipage}{0.33\textwidth}
		\vspace{0.7mm}
		\small
		\def\defaultHypSeparation{\hskip .2in}
		\begin{prooftree}
			\AxiomC{$\sem{b}_\sigma = \mathit{true}$}
			\UnaryInfC{$\langle \myif(b, \prog, \progg), \sigma \rangle \xrightarrow{\epsilon} \langle \prog, \sigma \rangle$}
		\end{prooftree}
	\end{minipage}%
	\begin{minipage}{0.33\textwidth}
		\vspace{3.9mm}
		\small
		\begin{prooftree}
			\AxiomC{}
			\UnaryInfC{$\langle \myskip\mysemi\prog, \sigma \rangle \xrightarrow{\epsilon} \langle \prog,  \sigma \rangle$}
		\end{prooftree}
	\end{minipage}
	\begin{minipage}{0.33\textwidth}
		\small
		\begin{prooftree}
			\AxiomC{$\langle \prog, \sigma \rangle \xrightarrow{\mu} \langle \prog', \sigma' \rangle$}
			\UnaryInfC{$\langle \prog\mysemi\progg, \sigma \rangle \xrightarrow{\mu} \langle \prog' \mysemi \progg,  \sigma' \rangle$}
		\end{prooftree}
	\end{minipage}
	
	\caption{Selection of small-step reduction rules. We write $\sem{e}_\sigma \in \intSet$ and $\sem{b}_\sigma \in \bool$ for the value of expression $e$ and $b$ in store $\sigma$, respectively.}\label{fig:semantic-rules}
\end{figure}

\paragraph{Syntax-Guided Synthesis.}

A Syntax-Guided Synthesis (SyGuS) problem is a triple $\Xi = (\{\tilde{f}_1, \ldots, 
\tilde{f}_n\}, \varrho, \{G_1, \ldots, G_n\})$, where $\tilde{f}_1, \ldots, \tilde{f}_n$ are function symbols, $\varrho$ is an SMT constraint over the function symbols $\tilde{f}_1, \ldots, \tilde{f}_n$, and $G_1, \ldots, G_n$ are grammars \cite{AlurBJMRSSSTU13}.
A solution for $\Xi$ is a vector of terms $\vec{e} = (e_1, \ldots, e_n)$ such that each $e_i$ is generated by grammar $G_i$, and $\varrho[\tilde{f}_1 / e_1, \ldots, \tilde{f}_n / e_n]$ holds (i.e., we replace each function symbol $\tilde{f}_i$ with expression $e_i$). 

\begin{example}\label{ex:sygus}
	Consider the SyGuS problem $\Xi = (\{\tilde{f}\}, \varrho, \{G\})$, where 
	\begin{align*}
		\varrho &:= \forall x, y\ldot \tilde{f}(x, y) \geq x \land \tilde{f}(x, y) \geq y \land (\tilde{f}(x, y) = x \lor \tilde{f}(x, y) = y)\\[2mm]
		G &:= \begin{cases}
			\begin{aligned}
				I &\to x \mid y \mid 0 \mid 1 \mid I + I \mid I - I \mid \mathit{ite}(B, I, I)\\[1mm]
				B &\to B \land B \mid B \lor B \mid \neg B \mid I = I \mid I \leq I \mid I \geq I.
			\end{aligned}
		\end{cases}
	\end{align*}
	This SyGuS problem constrains $\tilde{f}$ to be the function that returns the maximum of its arguments, and the grammar admits arbitrary piece-wise linear functions. 
	A possible solution to $\Xi$ would be $\tilde{f}(x, y) := \mathit{ite}(x \leq y, y, x)$.\demo
\end{example}

\paragraph{HyperLTL.}

As the basic specification language for hyperproperties, we use HyperLTL, an extension of LTL with explicit quantification over execution traces \cite{ClarksonFKMRS14}.
Let $\traceVars = \{\pi_1, \ldots, \pi_n\}$ be a set of \emph{trace variables}.
For a trace variable $\pi_j \in \traceVars$, we define $X_{\pi_j} :=  \{x_{\pi_j} \mid x \in X\}$ as a set of indexed program variables and $\vec{X} := X_{\pi_1} \cup \cdots \cup X_{\pi_n}$.
We include predicates from an arbitrary first-order theory $\mathfrak{T}$ to reason about the infinite variable domains in programs (cf.~\cite{BeutnerF22}), and denote satisfaction in $\mathfrak{T}$ with $\models^\mathfrak{T}$.
We write $\calF_{\vec{X}}$ for the set of first-order predicates over variables $\vec{X}$.
HyperLTL formulas are generated by the following grammar:
\begin{align*}
	\varphi &:= \forall \pi \ldot \varphi \mid \exists \pi \ldot \varphi \mid \psi\\
	\psi &:= \theta \mid \psi \land \psi \mid \psi \lor \psi  \mid \ltlN \psi  \mid \psi \ltlW \psi
\end{align*}
where $\pi \in \traceVars$, $\theta \in \calF_{\vec{X}}$, and $\ltlN$ and $\ltlW$ are the \emph{next} and \emph{weak-until} operator, respectively.
W.l.o.g., we assume that all variables in $\traceVars$ occur in the prefix exactly once.
We use the usual derived constants and connectives $\mathit{true}, \mathit{false}, \rightarrow$, and $\leftrightarrow$.

\begin{remark}\label{rem:safety}
	We only allow negation within the atomic predicates, effectively ensuring that the LTL-like body denotes a \emph{safety property} \cite{KupfermanV99}.
	The reason for this is simple:
	In our program semantics, we specifically allow for both infinite and \emph{finite} executions.
	Our repair approach is thus applicable to reactive systems but also handles (classical) programs that terminate.
	By requiring that the body denotes a safety property, we can easily handle arbitrary combinations of finite and infinite executions.
	Note that our logic supports \emph{arbitrary} quantifier alternations, so we can still express hyperliveness properties such as GNI.	
	\demo
\end{remark}

Let $\traceset \subseteq \states^\star$ be a set of traces.
For $t \in \traceset$ and $i < |t|$, we write $t(i)$ for the $i$th store in $t$. 
A trace assignment is a partial mapping $\Pi : \traceVars \rightharpoonup \traceset$ from trace variables to traces. 
We write $\Pi_{(i)}$ for the assignment $\vec{X} \to \intSet$ given by $\Pi_{(i)}(x_\pi) := \Pi(\pi)(i)(x)$, i.e., the value of $x_\pi$, is the value of $x$ in the $i$th step on the trace bound to $\pi$. 
We define the semantics inductively as:
\begin{align*}
	\Pi, i &\models_\traceset \psi &\text{if } \quad&\exists \pi \in \traceVars\ldot |\Pi(\pi)| \leq i\\
	\Pi, i &\models_\traceset \theta &\text{if } \quad&\Pi_{(i)} \models^\mathfrak{T} \theta\\
	\Pi, i&\models_\traceset \psi_1 \land \psi_2 &\text{if } \quad &\Pi,i \models_\traceset \psi_1 \text{ and } \Pi, i \models_\traceset \psi_2\\
	\Pi, i&\models_\traceset \psi_1 \lor \psi_2 &\text{if } \quad &\Pi,i \models_\traceset \psi_1 \text{ or } \Pi, i \models_\traceset \psi_2\\
	\Pi, i &\models_\traceset \ltlN \psi &\text{if } \quad &\Pi, i+1 \models_\traceset \psi\\
	\Pi, i &\models_\traceset \psi_1 \ltlW \psi_2 &\text{if } \quad &\big(\exists j \geq i\ldot \Pi, j \models_\traceset \psi_2 \text{ and }  \forall i \leq k < j\ldot \Pi, k \models_\traceset \psi_1\big) \text{ or} \\
	&\quad\quad\quad\quad\quad\quad\quad\quad\quad\quad\quad\quad\big(\forall j \geq i\ldot \Pi, j \models_\traceset \psi_1\big) \span \span\\
	\Pi, i &\models_\traceset \exists \pi \ldot \varphi  &\text{if } \quad&\exists t \in \traceset  \ldot \Pi[\pi \mapsto t], i  \models_\traceset  \varphi\\
	\Pi, i &\models_\traceset \forall \pi \ldot \varphi  &\text{if } \quad&\forall t \in \traceset  \ldot \Pi[\pi \mapsto t], i  \models_\traceset  \varphi
\end{align*}
As we deal with safety formulas (cf.~\Cref{rem:safety}), we let $\Pi, i$ satisfy any formula $\psi$ as soon as we have moved past the length of the shortest trace in $\Pi$ (i.e., $\exists \pi \in \traceVars\ldot |\Pi(\pi)| \leq i$). 
A program $\prog$ satisfies $\varphi$, written $\prog \models \varphi$, if $\emptyset, 0 \models_{\traces(\prog)} \varphi$, where $\emptyset$ denotes the trace assignment with an empty domain.

\paragraph{NSA.}

A {\em nondeterministic safety automaton} (NSA) over alphabet $\Sigma$ is a tuple $\calA = (Q, Q_0, \delta)$, where $Q$ is a finite set of states, $Q_0 \subseteq Q$ is a set of initial states, and $\delta \subseteq Q \times \Sigma \times Q$ is a transition relation.
A {\em run} of $\calA$ on a word $u \in \Sigma^\star$ is a sequence $q_0q_1 \cdots \in 
Q^\star$ such that $q_0 \in Q_0$ and for every $i < |u|$, $(q_i, u(i), q_{i+1}) \in \delta$. 
We write $\calL(\calA) \subseteq \Sigma^\star$ for the set of words on which $\calA$ has \emph{some} run.

\section{Program Repair by Symbolic Execution}\label{sec:simple-repair}

In our repair setting, we are given a pair $(\prog, \varphi)$ such that $\prog \not\models \varphi$, and try to construct a repaired program $\progg$ with $\progg \models \varphi$. 
In particular, we repair w.r.t.~a \emph{formal specification} instead of a set of input-output examples. 
The reason for this lies within the nature of the properties we want to repair against: 
When repairing against trace properties (i.e., functional specifications), it is often intuitive to write input-output examples that test a program's functional behavior. 
In contrast, hyperproperties do not directly reason about concrete functional behavior but rather about the abstract relation between multiple computations. 
For example, information-flow properties such as non-interference can be applied to arbitrary programs; independent of the program's functional behavior.
Perhaps counter-intuitively, in our hyper-setting, formal specifications are thus often easier to construct than input-output examples.

\begin{figure}[!t]
	
	\vspace{-2mm}
	
	\begin{minipage}{0.5\textwidth}
		\small
				\def\ScoreOverhang{1pt}
		\begin{prooftree}
			\AxiomC{}
			\UnaryInfC{$\langle x \myassign e, \nu, \alpha, \beta \rangle \symTo \langle \myskip, \nu[x \mapsto \sem{e}_\nu], \alpha, \beta \rangle$}
		\end{prooftree}
	\end{minipage}%
	\begin{minipage}{0.5\textwidth}
		\vspace{0.4mm}
		\small
				\def\ScoreOverhang{1pt}
				\begin{prooftree}
					\AxiomC{}
					\UnaryInfC{$\langle \myobserve, \nu, \alpha, \beta \rangle  \symTo \langle \myskip,  \nu, \alpha, \beta \cdot \nu \rangle$}
				\end{prooftree}
	\end{minipage}
	
	\vspace{2mm}
	
	\begin{minipage}{0.45\textwidth}
		\small
				\def\ScoreOverhang{1pt}
				\begin{prooftree}
					\AxiomC{}
					\UnaryInfC{$\langle \myskip\mysemi\prog, \nu, \alpha, \beta \rangle \symTo \langle \prog,  \nu, \alpha, \beta \rangle$}
				\end{prooftree}
	\end{minipage}%
	\begin{minipage}{0.55\textwidth}
		\small
				\def\ScoreOverhang{1pt}
		\begin{prooftree}
			\AxiomC{}
			\UnaryInfC{$\langle \myif(b, \prog, \progg), \nu, \alpha, \beta \rangle \symTo \langle \prog, \nu, \alpha \land \sem{b}_\nu, \beta \rangle$}
		\end{prooftree}
	\end{minipage}
	
	\vspace{2mm}
	
	\begin{minipage}{0.45\textwidth}
				\def\ScoreOverhang{1pt}
				\begin{prooftree}
					\AxiomC{$\langle \prog, \nu, \alpha, \beta \rangle \symTo \langle \prog', \nu', \alpha', \beta' \rangle$}
					\UnaryInfC{$\langle \prog\mysemi\progg, \nu, \alpha , \beta\rangle\symTo \langle \prog' \mysemi \progg,  \nu', \alpha', \beta' \rangle$}
				\end{prooftree}
	\end{minipage}%
	\begin{minipage}{0.55\textwidth}
		\small
				\def\ScoreOverhang{1pt}
				\vspace{4mm}
		\begin{prooftree}
			\AxiomC{}
			\UnaryInfC{$\langle \myif(b, \prog, \progg), \nu, \alpha , \beta\rangle \symTo \langle \progg, \nu, \alpha \land \neg \sem{b}_\nu, \beta \rangle$}
		\end{prooftree}
	\end{minipage}%
	
	\vspace{2mm}
	
	\begin{minipage}{1\textwidth}
		\small
				\def\ScoreOverhang{1pt}
		\begin{prooftree}
			\AxiomC{}
			\UnaryInfC{$\langle \mywhile(b, \prog), \nu, \alpha, \beta \rangle \symTo \langle \prog \mysemi \mywhile(b, \prog), \nu, \alpha \land \sem{b}_\nu,  \beta  \rangle$}
		\end{prooftree}
	\end{minipage}
	
	\vspace{2mm}
	
	\begin{minipage}{1\textwidth}
		\small
				\def\ScoreOverhang{1pt}
		\begin{prooftree}
			\AxiomC{}
			\UnaryInfC{$\langle \mywhile(b, \prog), \nu, \alpha, \beta \rangle \symTo \langle \myskip,  \nu, \alpha \land \neg \sem{b}_\nu, \beta \rangle$}
		\end{prooftree}
	\end{minipage}
	
	\caption{Small-step reduction rules for symbolic execution.}\label{fig:se}
\end{figure}

\subsection{Symbolic Execution}

The first step in our repair pipeline is the computation of a mathematical summary of (parts of) the program's executions using symbolic execution (SE) \cite{King76}.
In SE, we execute the program using symbolic placeholders instead of concrete values for variables, and explore all symbolic paths of a program (recording conditions that a concrete store needs to satisfy to take any given branch).
A \emph{symbolic store} is a function $\nu : X \to \expInt$ that maps each variable to an expression, and we write $\symstates := \{\nu \mid \nu : X \to \expInt\}$ for  the set of all symbolic stores.
A \emph{symbolic configuration} is then a tuple $\langle \prog, \nu, \alpha, \beta \rangle$, where 
$\prog$ is a program, $\nu \in \symstates$ is a symbolic store, $\alpha \in \calF_X$ is a first-order
formula over $X$ that records which conditions the current path should satisfy (called the 
\emph{path condition}), and $\beta \in \symstates^*$ is a sequence of symbolic stores recording the 
observations. 
For $e \in \expInt$ and $\nu \in \symstates$, we write $\sem{e}_\nu$ for the expression obtained by replacing each variable $x$ in $e$ with $\nu(x)$. 
For example, if $\nu = [x \mapsto x - 1, y \mapsto z * y]$, we have $\sem{x + y}_\nu = (x - 1) + (z * y)$. 
We give the symbolic execution relation $\symTo$ in \Cref{fig:se}.
We start the symbolic execution in symbolic store $\nu_0 := \big[ x \mapsto x \big]_{x \in X}$ that maps each variable to itself, path condition $\alpha_0 := \mathit{true}$, and an empty observation sequence $\beta_0 := \epsilon$.
Given a program $\prog$, a symbolic execution is a \emph{finite} sequence of symbolic configurations
\begin{align}\label{eq:symExec}
	\rho = \langle \prog,  \nu_0, \alpha_0, \beta_0 \rangle \symTo \langle \prog_1,  \nu_1, \alpha_1, \beta_1 \rangle \symTo \cdots \symTo \langle \prog_m, \nu_m, \alpha_m, \beta_m \rangle
\end{align}
We say execution $\rho$ is \emph{maximal} if $\prog_m = \myskip$, i.e., we cannot perform any more execution steps.
Given a symbolic execution $\rho$, we are interested in the path condition $\alpha_m$ (to ensure that we follow an actual program path), and the observation sequence $\beta_m$ (to evaluate the HyperLTL property).
We define a \emph{symbolic path} as a pair in $\calF_X \times  \symstates^*$, recording the path condition and symbolic observation sequence.
Each execution $\rho$ of the form in (\ref{eq:symExec}), yields a symbolic path $(\alpha_m, \beta_m)$. 
We call the symbolic path $(\alpha_m, \beta_m)$ \emph{maximal} if $\rho$ is maximal, and \emph{satisfiable} if $\alpha_m$ is satisfiable (i.e., some actual program execution can take a path summarized by $\rho$).
We write $\sympaths(\prog) \subseteq \calF_X \times  \symstates^*$ for the set of all satisfiable symbolic paths of $\prog$ and $\sympaths_\mathit{max}(\prog) \subseteq \calF_X \times  \symstates^*$ for the set of all satisfiable maximal symbolic paths.

\begin{remark}
	An interesting class of programs are those that are terminating and where $\sympaths_\mathit{max}(\prog)$ is \emph{finite}.
	This is either the case when the program is loop-free or has some upper bound on the number of loop executions (and thus control paths).
	Crucially, if $\sympaths_\mathit{max}(\prog)$ is finite, it provides a precise and complete mathematical summary of the program's executions.
	\demo
\end{remark}

\subsection{Symbolic Paths and Safety Automata}\label{sub:acc}

\begin{figure}[!t]
	\begin{encodingBox}{$\mathit{acc}^\psi_\Delta$}
		For $\Delta : \{\pi_1, \ldots, \pi_n\} \to  \symstates^*$, define $\mathit{acc}^\psi_\Delta := \bigvee_{q \in Q_{0, \psi}} \mathit{acc}_\Delta^{q, 0}$, \\
		where $\mathit{acc}_\Delta^{q, i} := \mathit{true}$ iff $\exists \pi\ldot |\Delta(\pi)| \leq i$, and otherwise
		\begin{align*}
			\mathit{acc}_\Delta^{q, i} := \bigvee_{(q, \iota, q') \in \delta_\psi}\bigg(\mathit{acc}_\Delta^{q', i+1}\, \land &\bigwedge_{\theta \in F \mid \theta \in \iota} \theta\Big[x_{\pi_j} \big/ \big(\Delta(\pi_j)(i)(x)\big[y / y_{\pi_j}\big]  \big) \Big] \;  \land \\
			&\bigwedge_{\theta \in F \mid \theta \not\in \iota} \neg \theta\Big[x_{\pi_j} \big/ \big(\Delta(\pi_j)(i)(x)\big[y / y_{\pi_j}\big]  \big) \Big]  \bigg)
		\end{align*}
		\vspace{-4mm}
	\end{encodingBox}
	\caption{Encoding for acceptance of $\psi$.}\label{fig:acc}
\end{figure}

We can use symbolic paths to approximate the HyperLTL 
semantics by explicitly considering path combinations.
Let $\varphi = \quant_1 \pi_1 \ldots \quant_n \pi_n \ldot \psi$ be a fixed HyperLTL formula, where 
$\quant_1, \ldots, \quant_n \in \{\forall, \exists\}$ are quantifiers, and $\psi$ is the LTL body of $\varphi$. 
Further, let $F \subseteq \calF_{\vec{X}}$ be the \emph{finite} set of predicates used in $\psi$.
Due to our syntactic safety restriction on LTL formulas, we can construct an NSA $\calA_\psi = 
(Q_\psi, Q_{0, \psi}, \delta_\psi)$ over alphabet $2^F$ accepting exactly the words that satisfy $\psi$ \cite{KupfermanV99}.

Assume $\Delta : \{\pi_1, \ldots, \pi_n\} \to  \symstates^*$ is a function that assigns each path variable $\pi_1, 
\ldots, \pi_n$ a symbolic observation sequence. 
We design a formula $\mathit{acc}^{\psi}_\Delta$, which encodes that the symbolic observation sequences in $\Delta$ have an accepting prefix in $\calA_\psi$, given in \Cref{fig:acc}.
The intermediate formula $\mathit{acc}_\Delta^{q, i}$ encodes that the observations in $\Delta$ have some run from state $q$ in the $i$th step. 
For all steps $i$, longer than the shortest trace in $\Delta$, we accept (i.e., $\mathit{acc}_\Delta^{q, i} := \mathit{true}$, similar to our HyperLTL semantics).
Otherwise, we require some transition $(q, \iota, q') \in \delta_\psi$ such that $\mathit{acc}_\Delta^{q', i+1}$ holds, and the label $\iota \in 2^F$ holds in step $i$. 
To encode the latter, we use the symbolic observation sequences in $\Delta$:
For every predicate $\theta \in F$, we require that $\theta \in \iota$ iff 
$\theta\big[x_{\pi_j} \big/ \big(\Delta(\pi_j)(i)(x)[y / y_{\pi_j}]\big) \big]$.
That is, we replace variable $x_{\pi_j}$ with the expression $\Delta(\pi_j)(i)(x)[y / y_{\pi_j}]$, i.e., we look up the expression bound to variable $x$ in the $i$th step on $\Delta(\pi_j)$, and -- within this expression -- index all variables with $\pi_j$ (i.e., replace each variable $y \in X$ with $y_{\pi_j} \in X_{\pi_j}$).

\subsection{Encoding for HyperLTL}\label{sub:simple-repair-encoding}

\begin{figure}[!t]
	\begin{encodingBox}{$\mathit{enc}^\varphi_\symPathSet$}
		For $\symPathSet \subseteq \calF_X \times \symstates^*$, define $\mathit{enc}^\varphi_\symPathSet := \mathit{enc}^\varphi_{\symPathSet, \emptyset}$, where
			\begin{align*}
				\mathit{enc}_{\symPathSet, \Delta}^{\psi} &:= \mathit{acc}^\psi_\Delta\\
			\mathit{enc}_{\symPathSet, \Delta}^{\exists \pi_j\ldot \varphi'} &:= \bigexists_{x_{\pi_j} \in X_{\pi_j}} x_{\pi_j}\ldot \bigvee_{(\alpha, \beta) \in \symPathSet} \Big( \alpha[x / x_{\pi_j}] \land  	\mathit{enc}_{\symPathSet, \Delta[\pi_j \mapsto \beta]}^{\varphi'} \Big)  \\
			\mathit{enc}_{\symPathSet, \Delta}^{\forall \pi_j\ldot \varphi'} &:= \bigforall_{x_{\pi_j} \in X_{\pi_j}} x_{\pi_j}\ldot \bigwedge_{(\alpha, \beta) \in \symPathSet} \Big( \alpha[x / x_{\pi_j}] \rightarrow \mathit{enc}_{\symPathSet, \Delta[\pi_j \mapsto \beta]}^{\varphi'}  \Big)
		\end{align*}
	\end{encodingBox}
	
	\caption{Encoding of the HyperLTL semantics on symbolic paths $\symPathSet$. }\label{fig:enc}
\end{figure}

Let $\symPathSet \subseteq \calF_X \times \symstates^*$ be a finite 
set of symbolic paths and consider the formula $\mathit{enc}^\varphi_\symPathSet$ in \Cref{fig:enc}.
Intuitively, the formula encodes the satisfaction of $\varphi$ on the symbolic paths in $\symPathSet$. 
For this, we maintain a \emph{partial} mapping $\Delta : \{\pi_1, \ldots, \pi_n\} \rightharpoonup  \symstates^*$, and for each subformula $\varphi'$ we define an intermediate formula $\mathit{enc}_{\symPathSet, \Delta}^{\varphi'}$.
If we reach the LTL body $\psi$, we define $\mathit{enc}_{\symPathSet, \Delta}^{\psi} := \mathit{acc}^\psi_\Delta$, stating that the symbolic observation sequences in $\Delta$ satisfy $\psi$ (cf.~\Cref{fig:acc}).
Each trace quantifier is then resolved on the symbolic paths in $\symPathSet$.
Concretely, for a subformula $\exists \pi_j \ldot \varphi'$, we existentially quantify over variables $X_{\pi_j}$ and \emph{disjunctively} pick a symbolic path $(\alpha, \beta) \in \symPathSet$.
We require that path condition $\alpha$ holds (after replacing each variable $x$ with $x_{\pi_j}$), and that the remaining formula $\varphi'$ is satisfied if we bind observation sequence $\beta$ to $\pi_j$ (i.e., $\mathit{enc}_{\symPathSet, \Delta[\pi_j \mapsto \beta]}^{\varphi'}$). 

\begin{proposition}
	If $\progg$ is a terminating program and $\sympaths_\mathit{max}(\progg)$ is finite, then $\progg \models \varphi$ if and only if $\mathit{enc}^\varphi_{\sympaths_\mathit{max}(\progg)}$. 
\end{proposition}

The above proposition essentially states that we can use SE to verify a program (with finitely-many symbolic paths) against HyperLTL formulas with \emph{arbitrary} quantifier alternations. 
This is in sharp contrast to existing SE-based approaches, which only apply to $k$-safety properties (i.e., $\forall^*$ HyperLTL formulas) \cite{FarinaCG19,DanielBR20,TsoupidiBB21,TiraboschiRR23,DanielBR21}.
To the best of our knowledge, ours is the first approach that can check properties containing arbitrary alternations on fragments of infinite-state systems. 
Previous methods focus on finite-state systems \cite{HsuSB21,HsuSSB23,BeutnerF22b,BeutnerF23,FinkbeinerRS15,CoenenFST19,BeutnerF24} or only consider restricted quantifier structures \cite{FarzanV19,BeutnerF22,Beutner24,ShemerGSV19,UnnoTK21,ItzhakySV24}. 

\paragraph{Alternation-Free Formulas. }

In many situations, we cannot explore \emph{all} symbolic paths of a program $\progg$ (i.e., $\sympaths_\mathit{max}(\progg)$ is infinite).
However, even by just exploring a subset of paths, our encoding still allows us to draw conclusions about the full program as long as the formula is \emph{alternation-free}.

\begin{proposition}\label{prop:existential}
	Assume $\varphi$ is a $\exists^*$ HyperLTL formula and $\symPathSet \subseteq \sympaths_\mathit{max}(\progg)$ is a finite set of maximal symbolic paths.
	If $\mathit{enc}^\varphi_{\symPathSet}$, then $\progg \models \varphi$. 
\end{proposition}

\begin{proposition}\label{prop:universal}
	Assume $\varphi$ is a $\forall^*$ HyperLTL formula and $\symPathSet \subseteq \sympaths(\progg)$ is a finite set of (not necessarily maximal) symbolic paths.
	If $\neg \mathit{enc}^\varphi_{\symPathSet}$, then $\progg \not\models \varphi$. 
\end{proposition}

In particular, we can use \Cref{prop:universal} for our repair approach for $\forall^*$ properties (which captures many properties of interest, such as non-interference, cf.~\ref{eq:edas}).
If we symbolically execute a program to some fixed depth (and thus capture a subset of the symbolic paths), any possible repair must satisfy the bounded property described in $\mathit{enc}^\varphi_{\symPathSet}$ (cf.~\Cref{sub:add-functions}). 
Note that this does not ensure that the repair patch that fulfills $\mathit{enc}^\varphi_{\symPathSet}$ is correct on the entire program;  $\mathit{enc}^\varphi_{\symPathSet}$ merely describes a \emph{necessary} condition any possible repair needs to satisfy.
In our experiments (cf.~\Cref{sec:implementation}), we (empirically) found that the repair for the bounded version also serves as a repair for the full program in many instances.

\subsection{Program Repair using SyGuS}\label{sub:add-functions}

Using SE and our encoding, we can now outline our basic SyGuS-based repair approach.
Assume $\prog \not\models \varphi$ is the program that should be repaired.  
As in other semantic-analysis-based repair frameworks \cite{MechtaevYR16,NguyenQRC13}, we begin our repair by predicting fault locations \cite{WongGLAW16}
within the program, i.e., locations that are likely to be responsible for the violation of $\varphi$.  
In our later experiments, we assume that these locations are provided by the user.
After we have identified a set of $n$ repair locations, we instrument $\prog$ by replacing the 
expressions in all repair locations with fresh {\em function symbols}. 
That is, if we want to repair statement $x \myassign e$, $\myif(b, \prog_1, \prog_2)$, or $\mywhile(b, \prog)$, we replace the statement with $x \myassign \tilde{f}(x_1, \ldots, x_m)$, $\myif(\tilde{f}(x_1, \ldots, x_m), \prog_1, \prog_2)$, or $\mywhile(\tilde{f}(x_1, \ldots, x_m), \prog)$, respectively, for some fresh function symbol $\tilde{f}$ and program variables $x_1, \ldots, x_m \in X$ (inferred using a lightweight dependency analysis).  
Let $\progg$ be the resulting program, which contains function symbols, $\tilde{f}_1, \ldots, \tilde{f}_n$. 
We symbolically execute $\progg$, leading to a set of symbolic paths $\symPathSet$ containing $\tilde{f}_1, \ldots, \tilde{f}_n$, and define the SyGuS problem $\Xi_\symPathSet := (\{\tilde{f}_1, \ldots,\tilde{f}_n\}, \mathit{enc}^\varphi_{\symPathSet}, \{G_1, \ldots, G_n\})$.
Here, we fix a grammar $G_i$ for each function symbol $\tilde{f}_i$, based on the type and context of each repair location. 
Note that $\mathit{enc}^\varphi_{\symPathSet}$ now constitutes an SMT constraint over $\tilde{f}_1, \ldots, \tilde{f}_n$. 
Any solution for $\Xi_\symPathSet$ thus defines concrete expressions for $\tilde{f}_1, \ldots, \tilde{f}_n$ such that the symbolic paths in $\symPathSet$ satisfy $\varphi$.
Concretely, let $\vec{e} = (e_1, \ldots, e_n)$ be a solution to $\Xi_\symPathSet$. 
Define $\progg[\vec{e}] :=\progg[\tilde{f}_1 / e_1, \ldots, \tilde{f}_n/ e_n]$, i.e., we replace each function symbol $\tilde{f}_i$ by expression $e_i$.
As $\vec{e}$ is a solution to $\Xi_\symPathSet$, we directly obtain that  $\progg[\vec{e}] $ satisfies $\varphi$; at least restricted to the executions captured by the symbolic paths in $\symPathSet$. 
Afterward, we can \emph{verify} that  $\progg[\vec{e}] $ indeed satisfies $\varphi$ (even on paths not explored in $\symPathSet$), using existing hyperproperty verification techniques \cite{FarzanV19,BeutnerF22,ShemerGSV19,ItzhakySV24,UnnoTK21}.

\begin{example}\label{ex:edas-trace}
	Consider the EDAS program $\prog$ in \Cref{fig:edas}, and let $\progg$ be the modified program where the assignment in line \ref{line:repair-motivations} is replaced with a fresh function symbol $\tilde{f}$.
	Define $X := \{\exm{phase}, \exm{title}, \exm{session}, \exm{decision}\}$. 
	If we perform SE on $\progg$, we get two symbolic paths $\symPathSet_\progg = \{(\alpha_1, \beta_1),  (\alpha_2, \beta_2) \}$, where $\alpha_1 = (\tilde{f}(X) = \exm{"Accept"})$, $\alpha_2 = (\tilde{f}(X) \neq \exm{"Accept"})$, $\beta_1 = \big[ [\ldots], [\exm{print} \mapsto \exm{title + session}, \exm{decision} \mapsto \tilde{f}(X), \ldots]   \big]$, and $\beta_2 = \big[ [\ldots], [\exm{print} \mapsto \exm{title}, \allowbreak \exm{decision} \mapsto \tilde{f}(X), \ldots]   \big]$. 
	For illustration, we consider the simple trace property $\varphi_\mathit{trace} = \forall \pi\ldot \ltlN (\exm{print}_\pi = \exm{title}_\pi)$. 
	If we construct $\mathit{enc}_{\symPathSet_\progg}^{\varphi_\mathit{trace}}$, we get
	\begin{align*}
		\bigforall_{x_\pi \in X_\pi} x_\pi\ldot \, &\Big( \tilde{f}(X_\pi) = \exm{"Accept"} \to  \exm{title$_\pi$ + session$_\pi$} = \exm{title}_\pi \Big) \; \land \\
		&\Big( \tilde{f}(X_\pi) \neq \exm{"Accept"} \to  \exm{title}_\pi = \exm{title}_\pi  \Big),
	\end{align*}
	allowing the simple SyGuS solution $\tilde{f}(X_\pi) := \exm{"Reject"}$. \demo
\end{example}

\section{Transparent Repair}\label{sec:trans-repair}

As argued in \Cref{sec:overview}, searching for \emph{any} repair (as in \Cref{sec:simple-repair}) often returns a patch that severely changes the functional behavior of the program. 
In this paper, we study a principled constraint-based approach on how to guide the search towards a useful repair without requiring extensive additional specifications. 
Our method is based on the simple idea that the repair should be somewhat \emph{close} to the original program.
Crucially, we define ``closeness'' via \emph{rigorous} systems of (SyGuS) constraints, guiding our constraint-based repair towards minimal patches, with \emph{guaranteed} quality.
In this section, we introduce the concept of a (fully) transparent repair.
In \Cref{sec:iterative}, we adapt this idea and present a more practical adaption in the form of iterative repair.

\subsection{Transparency}

Our transparent repair approach is motivated by ideas from the \emph{enforcement} literature \cite{NgoMMP15}.
In enforcement, we do not repair the program (i.e., we do not manipulate its source code) but rather let an enforcer run alongside the program and intervene on unsafe behavior (by, e.g., overwriting the output).
The obvious enforcement strategy would thus always intervene, effectively overwriting all program behaviors with some dummy (but safe) behavior.
To avoid such trivial enforcement, researchers have developed the notion of \emph{transparency} (also called \emph{precision} \cite{NgoMMP15}). 
Transparency states that the enforcer should not intervene unless an intervention is \emph{absolutely necessary} to satisfy the safety specification, i.e., a safe prefix of the program execution should never trigger the enforcer.

\paragraph{Transparent Repair.}

The original transparency definition is specific to program enforcement and refers to the \emph{time step} in which the enforcer intervenes. 
We propose an adoption to the repair setting based on the idea of preserving as much input-output behavior of the original program as possible. 
Let $X_\mathit{out} \subseteq X$ be a set of program variables defining the output. 
For two stores $\sigma, \sigma' \in \states$, we write $\sigma \neq_{X_\mathit{out}} \! \sigma'$ if $\sigma(x) \neq \sigma'(x)$ for some $x \in X_\mathit{out}$, and extend $\neq_{X_\mathit{out}}$  position-wise to sequences of stores. 

\begin{definition}[Fully Transparent Repair]\label{def:trans}
	Assume $\varphi = \forall \pi_1\ldots \forall \pi_n\ldot \psi$ is a $\forall^*$ HyperLTL formula and $\prog, \progg$ are programs.
	We say $\progg$ is a fully transparent repair of $(\prog, \varphi)$, if {(1)} $\progg \models \varphi$, and {(2)} for every store $\sigma \in \states$ where $\obs(\prog, \sigma) \neq_{X_\mathit{out}} \! \obs(\progg, \sigma)$, there exist stores $\sigma_1, \ldots , \sigma_n \in \states$ such that $\big[ \pi_j \mapsto \obs(\prog, \sigma_j) \big]_{j = 1}^n, 0 \not\models \psi$, and $\sigma = \sigma_j$ for some $1 \leq j \leq n$. 
\end{definition}

Our definition reasons about inputs $\sigma$ on which the output behavior of $\progg$ differs from the original program $\prog$. 
Any such input $\sigma$ must take part in a violation of $\varphi$ on the original program $\prog$. 
Phrased differently, the repair may only change $\prog$'s behavior on executions that take part in a combination of $n$ traces that violate $\varphi$.
Note that, similar to enforcement approaches \cite{CoenenFHHS21,NgoMMP15}, our transparency definition only applies to $\forall^*$ formulas.
As soon as the property includes existential quantification, we can no longer formalize when some execution is ``part of a violation of $\varphi$''.
We will extend the central idea underpinning transparency to arbitrary HyperLTL formulas in \Cref{sec:iterative}.

\subsection{Encoding for Transparent Repair}
\label{sub:trans-encoding}

\begin{figure}[!t]
	\begin{encodingBox}{$\mathit{trans}^\varphi_{\symPathSet_\prog, \symPathSet_\progg}$}
		For $\symPathSet_\prog, \symPathSet_\progg \subseteq \calF_X \times \symstates^*$, define $\mathit{trans}^\varphi_{\symPathSet_\prog, \symPathSet_\progg}$ as
		\begin{align*}
			&\bigforall_{x \in X} x\ldot \Bigg(  \bigg[ \bigvee_{(\alpha_\prog, \beta_\prog) \in \symPathSet_\prog} \bigvee_{(\alpha_\progg, \beta_\progg) \in \symPathSet_\progg}  \alpha_\prog \land \alpha_\progg \land \!\!\! \bigvee_{i = 0}^{\min(|\beta_\prog|, |\beta_\progg|) - 1} \!\!\!\!\bigvee_{x \in X_\mathit{out}} \beta_\prog(i)(x) \neq \beta_\progg(i)(x) \bigg] \to \\
			&\quad\quad  \bigexists_{x_{\pi_1} \in X_{\pi_1}} \!\!\!\!\! x_{\pi_1} \ldot \cdots \bigexists_{x_{\pi_n} \in X_{\pi_n}} \!\!\!\!\! x_{\pi_n} \ldot \bigvee_{(\alpha_{\pi_1}, \beta_{\pi_1}) \in \symPathSet_\prog} \cdots \bigvee_{(\alpha_{\pi_n}, \beta_{\pi_n}) \in \symPathSet_\prog}  \\
			&\quad\quad \quad\quad \Big (\bigwedge_{j=1}^n \alpha_{\pi_j}[x/ x_{\pi_j}] \Big) \land \Big(\bigvee_{j = 1}^n \bigwedge_{x \in X} x = x_{\pi_j}\Big) \land \neg \mathit{acc}^\psi_{[\pi_1 \mapsto \beta_{\pi_1}, \ldots, \pi_n \mapsto \beta_{\pi_n}]} \Bigg)
		\end{align*} 
	\end{encodingBox}
	\caption{Encoding for (fully) transparent repair.}\label{fig:trans}
\end{figure}

Given two finite sets of symbolic paths $\symPathSet_\prog, \symPathSet_\progg \subseteq \calF_X \times \symstates^*$, we define formula $\mathit{trans}^\varphi_{\symPathSet_\prog, \symPathSet_\progg}$ in \Cref{fig:trans}.
The premise states that $X$ defines some input on which $\prog$ and $\progg$ differ in their output.
That is, for some symbolic paths $(\alpha_\prog, \beta_\prog) \in \symPathSet_\prog$ and $(\alpha_\progg, \beta_\progg)\in \symPathSet_\progg$, the path conditions $\alpha_\prog$ and $\alpha_\progg$ hold, but the symbolic observation sequences yield some different values for some $x \in X_\mathit{out}$.
In this case, we require that there exist $n$ symbolic paths $(\alpha_{\pi_1}, \beta_{\pi_1}), \ldots, (\alpha_{\pi_n}, \beta_{\pi_n}) \in \symPathSet_\prog$ and concrete inputs $X_{\pi_1}, \ldots, X_{\pi_n}$, such that 
{(1)} the path conditions $\alpha_{\pi_1}, \ldots, \alpha_{\pi_n}$ hold; 
{(2)} the assignment to some $X_{\pi_j}$ equals $X$; and 
{(3)} the symbolic observation sequences $\beta_{\pi_1}, \ldots, \beta_{\pi_n}$ violate $\psi$ (cf.~\Cref{fig:acc}).

\begin{proposition}
	If $\prog, \progg$ are terminating and $\sympaths_\mathit{max}(\prog), \sympaths_\mathit{max}(\progg)$ are finite, then $\progg$ is a fully transparent repair of $(\prog, \varphi)$ if and only if
	\begin{align*}
		\mathit{enc}_{\sympaths_\mathit{max}(\progg)}^\varphi \land \mathit{trans}^\varphi_{\sympaths_\mathit{max}(\prog), \sympaths_\mathit{max}(\progg)}.
	\end{align*}
\end{proposition}

\begin{example}\label{ex:edas-trace-trans}
	We illustrate transparent repairs using \Cref{ex:edas-trace}.
	If we set $X_\mathit{out} := \{\exm{decision}\}$, and compute $\mathit{trans}^{\varphi_\mathit{trace}}_{\symPathSet_\prog, \symPathSet_\progg}$, we get
	\begin{align*}
		&\bigforall_{x \in X} x \ldot \; \big(\exm{decision} \neq \tilde{f}(X) \big) \to \Big( \big( \exm{decision} = \exm{"Accept"} \; \land \\[-1mm]
		&\quad\quad\exm{title + session}  \neq \exm{title} \big) \lor \big(\exm{decision} \neq \exm{"Accept"}\land \exm{title} \neq \exm{title} \big)\Big).
	\end{align*}
	For simplicity, we directly resolved the existentially quantified variables $X_{\pi}$ with $X$ and summarized all path constraints in the premise.
	The na\"ive solution $\tilde{f}(X) := \exm{"Reject"}$ from \Cref{ex:edas-trace} no longer satisfies $\mathit{trans}^{\varphi_\mathit{trace}}_{\symPathSet_\prog, \symPathSet_\progg}$.
	Instead, a possible SyGuS solution for $\mathit{enc}_{\symPathSet_\progg}^{\varphi_\mathit{trace}} \land  \mathit{trans}^{\varphi_\mathit{trace}}_{\symPathSet_\prog, \symPathSet_\progg}$ is
	\begin{align*}
		\tilde{f}(X) := \mathit{ite}\big(\exm{decision} = \exm{"Accept"} \land \exm{session} \neq \exm{""}, \exm{"Reject"}, \exm{decision}\big).
	\end{align*}
	This solution only changes the decision if the \ex{decision} is \exm{"Accept"} \emph{and} the \ex{session} does not equal the empty string, i.e., it changes the program's \ex{decision} on exactly those traces that violate $\varphi_\mathit{trace} = \forall \pi\ldot \ltlN (\exm{print}_\pi = \exm{title}_\pi)$.
	\demo
\end{example}

\section{Iterative Repair}
\label{sec:iterative}

Our full transparency definition only applies to $\forall^*$ properties, and, even on $\forall^*$ formulas, might yield undesirable results:
In some instances, \Cref{def:trans} limits which traces may be changed by a repair, 
potentially resulting in the absence of any repair.
In other instances (including the EDAS example), many paths (in the EDAS example, \emph{all} paths) take part in \emph{some} violation of the hyperproperty, so full transparency does not impose any additional constraints.
In the EDAS example, this would again allow the na\"ive repair \exm{decision = "Reject"}.
To alleviate this, we introduce an \emph{iterative repair} approach that follows the same philosophical principle as (full) transparency (i.e., search for repairs that are close to the original program), but allows for the \emph{iterative} discovery of better and better repair patches.

\begin{definition}\label{def:iter}
	Assume $\varphi$ is a HyperLTL formula and $\prog$, $\progg$, and $\proggg$ are programs. 
	We say repair $\progg$ is a better repair than $\proggg$ w.r.t.~$(\prog, \varphi)$ if {(1)} $\progg 
	\models \varphi$, {(2)} for every $\sigma \in \states$, where $\obs(\prog, \sigma) 
	\neq_{X_\mathit{out}} \! \obs(\progg, \sigma)$, we have $\obs(\prog, \sigma) \neq_{X_\mathit{out}}  \!
	\obs(\proggg, \sigma)$, and {(3)} for some $\sigma \in \states$, we have 
	$\obs(\prog, \sigma) \neq_{X_\mathit{out}} \!\! \obs(\proggg, \sigma)$ but $\obs(\prog, \sigma) =_{X_\mathit{out}} \! \obs(\progg, \sigma)$.
\end{definition}

Intuitively, $\progg$ is better than $\proggg$ if it preserves at least all those behaviors of $\prog$ already preserved by $\proggg$, i.e., $\progg$ is only allowed to deviate from $\prog$ on inputs where $\proggg$ already deviates.
Moreover, it must be strictly better than $\proggg$, i.e., preserve at least one additional behavior.

\begin{figure}[!t]
	\begin{encodingBox}{$\mathit{iter}_{\symPathSet_\prog, \symPathSet_\proggg, \symPathSet_{\progg}}$}
		For $\symPathSet_\prog, \symPathSet_\proggg, \symPathSet_\progg \subseteq \calF_X \times \symstates^*$, define $\mathit{iter}_{\symPathSet_\prog, \symPathSet_\proggg, \symPathSet_{\progg}}$ as
		\begin{align*}
			&\Bigg(\bigforall_{x \in X} x \ldot \bigwedge_{(\alpha_\prog, \beta_\prog) \in \symPathSet_\prog} \bigwedge_{(\alpha_\proggg, \beta_\proggg) \in \symPathSet_\proggg} \bigwedge_{(\alpha_\progg, \beta_\progg) \in \symPathSet_{\progg}} \\
			&\quad\quad\quad\bigg[ \alpha_\prog \land \alpha_\proggg\land  \alpha_\progg \land \bigvee_{i = 0}^{\min(|\beta_\prog|, |\beta_{\progg}|)-1} \!\! \bigvee_{x \in X_\mathit{out}} \obs_\prog(i)(x)  \neq \obs_{\progg}(i)(x) \bigg] \to \\
			&\quad\quad\quad\bigg[\bigvee_{i = 0}^{\min(|\beta_\prog|, |\beta_{\proggg}|)-1} \!\! \bigvee_{x \in X_\mathit{out}} \obs_\prog(i)(x)  \neq \obs_{\proggg}(i)(x) \bigg] \Bigg) \, \land \\
			&\Bigg(\bigexists_{x \in X} x\ldot \bigvee_{(\alpha_\prog, \beta_\prog) \in \symPathSet_\progg} \bigvee_{(\alpha_\proggg, \beta_\proggg) \in \symPathSet_\proggg} \bigvee_{(\alpha_\progg, \beta_\progg) \in \symPathSet_\progg}  \alpha_\prog \land  \alpha_\proggg \land  \alpha_\progg \; \land \\
			&\quad\quad\quad\bigg[\bigvee_{i = 0}^{\min(|\beta_{\prog}|, |\beta_\proggg|) - 1} \!\! \bigvee_{x \in X_\mathit{out}} \obs_{\prog}(i)(x)  \neq \obs_\proggg(i)(x) \bigg] \, \land \\
			&\quad\quad\quad\bigg[\bigwedge_{i = 0}^{\min(|\beta_{\prog}|, |\beta_\progg|) - 1} \!\! \bigwedge_{x \in X_\mathit{out}} \obs_{\prog}(i)(x)  = \obs_\progg(i)(x) \bigg] \Bigg)
		\end{align*}
	\end{encodingBox}
	\caption{Encoding for iterative repair.}\label{fig:iter}
\end{figure}

\subsection{Encoding for Iterative Repair}

As before, we show that we can encode \Cref{def:iter} via a repair constraint.
Let $\symPathSet_\prog$, $\symPathSet_\proggg, \symPathSet_\progg \subseteq \calF_X \times \symstates^*$ be finite sets of symbolic paths, and define $\mathit{iter}_{\symPathSet_\prog, \symPathSet_\proggg, \symPathSet_{\progg}}$ as in \Cref{fig:iter}.

\begin{proposition}
	If $\prog$, $\progg$, and $\proggg$ are terminating programs and $\sympaths_\mathit{max}(\prog)$, $\sympaths_\mathit{max}(\proggg)$, and $\sympaths_\mathit{max}(\progg)$ are finite, then $\progg$ is a better repair than $\proggg$, w.r.t., $(\prog, \varphi)$ if and only if 
	\begin{align*}
		\mathit{enc}_{\sympaths_\mathit{max}(\progg)}^\varphi \land \mathit{iter}_{\sympaths_\mathit{max}(\prog), \sympaths_\mathit{max}(\proggg), \sympaths_\mathit{max}(\progg)}.
	\end{align*}
\end{proposition}

\subsection{Iterative Repair Loop}

We sketch our iterative repair algorithm in \Cref{alg:rep}. 
In line \ref{line:localize}, we infer the locations that we want to repair from user annotations. 
We leave the exploration of automated fault localization techniques specific for hyperproperties as future work, and, in our experiments, assume that the user marks potential repair locations.
In line \ref{line:uniterpret}, we instrument $\prog$ by replacing all repair locations in $\mathit{locs}$ with fresh function symbols. 
At the same time, we record the original expression at all those locations as a vector $\vec{e}_\prog$. 
Subsequently, we perform symbolic execution on the skeleton program $\progg$ (i.e., the program that contains fresh function symbols), yielding a set of symbolic paths $\symPathSet$ containing function symbols (line \ref{line:se}).  
Initially, we now search for \emph{some} repair of $\varphi$ by using the SyGuS constraint $\mathit{enc}_\symPathSet^\varphi$, giving us an initial repair patch in the form of some expression vector $\vec{e}$ (line \ref{line:sygus1}). 
Afterward, we try to iteratively improve upon the repair solution $\vec{e}$ found previously.
For this, we consider the SyGuS constraint  $\mathit{enc}_{\symPathSet}^\varphi \land \mathit{iter}_{\symPathSet[\vec{e}_\prog], \symPathSet[\vec{e}], \symPathSet}$ where we replaced each function symbol in $\symPathSet$ with $\vec{e}_\prog$ to get the symbolic paths of the original program (denoted $\symPathSet[\vec{e}_\prog]$), and with $\vec{e}$ to get the symbolic paths of the previous repair (denoted $\symPathSet[\vec{e}]$) (line \ref{line:sygus2}).
If this SyGuS constraint admits a solution $\vec{e}'$, we set $\vec{e}$ to $\vec{e}'$ and repeat with a further improvement iteration (line \ref{line:update}).
If the SyGuS constraint is unsatisfiable (or, e.g., a timeout is reached, or the number of iterations is bounded) (written $\vec{e}' = \bot$), we return the last solution we found, i.e., the program $\progg[\vec{e}]$ (line \ref{line:return}).
By using a single set of symbolic paths $\symPathSet$ of the skeleton program $\progg$, we can optimize our query construction. 
For example, in $\mathit{iter}_{\symPathSet[\vec{e}_\prog], \symPathSet[\vec{e}], \symPathSet}$, we consider all 3 tuples of symbolic paths leading to a potentially large SyGuS query. 
As we use a common set of paths $\symPathSet$ we can prune many path combinations.
For example, on fragments preceding a repair location, we never have to combine contradicting branch conditions.

\begin{wrapfigure}{R}{0.53\linewidth}
	\vspace{-15.5mm}
	\begin{minipage}{\linewidth}
		\begin{algorithm}[H]
			\caption{Iterative repair algorithm}\label{alg:rep}
			\vspace{-2mm}
\begin{code}
def iterativeRepair($\prog$,$\varphi$):
@@$\mathit{locs}$ := faultLocalization($\prog$,$\varphi$) (*\label{line:localize}*)
@@$\progg$,$\vec{e}_\prog$ := instrument($\prog$,$\mathit{locs}$) (*\label{line:uniterpret}*)
@@$\symPathSet$ := symbolicExecution($\progg$) (*\label{line:se}*)
@@$\vec{e}$ := SyGuS($\mathit{enc}^\varphi_{\symPathSet}$) (*\label{line:sygus1}*)
@@repeat: (*\label{line:loop}*)
@@@@$\vec{e}'$ := SyGuS($\mathit{enc}_{\symPathSet}^\varphi \land \mathit{iter}_{\symPathSet[\vec{e}_\prog], \symPathSet[\vec{e}], \symPathSet}$) (*\label{line:sygus2}*)
@@@@if ($\vec{e}'$ = $\bot$) then 
@@@@@@return $\progg[\vec{e}]$ (*\label{line:return}*) 
@@@@else 
@@@@@@$\vec{e}$ := $\vec{e}'$ (*\label{line:update}*)
\end{code}
\vspace{-2mm}
		\end{algorithm}
	\end{minipage}
	\vspace{-3mm}
\end{wrapfigure}
 
\section{Implementation and Evaluation}\label{sec:implementation}

We have implemented our repair techniques from \Cref{sec:simple-repair,sec:trans-repair,sec:iterative} 
in a proof-of-concept prototype called \tool{}, which takes as input a HyperLTL formula and a program in a 
minimalist {\sf \small C}-like language featuring Booleans, integers, and strings.
We use \texttt{spot}~\cite{Duret-LutzRCRAS22} to translate LTL formulas to NSAs.
\tool{} can use any solver supporting the SyGuS input format \cite{AlurBJMRSSSTU13}; we use \texttt{cvc5} (version 1.0.8) \cite{BarbosaBBKLMMMN22} as the default solver in all experiments.
In \tool{}, the user can determine what SyGuS grammar to use, guiding the solver towards a particular (potentially domain-specific) solution.
By default, \tool{} repairs integer and Boolean expressions using piece-wise linear functions (similar to \Cref{ex:sygus}), and string-valued expressions by a grammar allowing selected string constants and concatenation of string variables. 
All results in this paper were obtained using a Docker container of \tool{} running on an Apple M1 Pro CPU and 32 GB of memory.

\paragraph{Scalability Limitations.}

As we repair for hyperproperties, we necessarily need to reason about the combination of paths, requiring us to analyze multiple paths simultaneously. 
Unsurprisingly, this limits the scalability of our repair.
Consequently, we cannot tackle programs with hundreds of LoC, where existing (\emph{functional}) 
APR approaches collect a small summary that only depends on the number of input-output examples 
(see, e.g., \emph{angelic forests} \cite{MechtaevYR16}).
However, our experiments with \tool{} attest that -- while we can only handle small programs -- our 
approach can find \emph{complex} repair solutions that go beyond previous repair 
approaches for hyperproperties (cf.~\Cref{sec:related-work}).

\subsection{Iterative Repair for Hyperproperties}
\label{sub:iterative}

\begin{wraptable}{R}{0.54\linewidth}
	\vspace{-8mm}
	\caption{We depict the number of improvement iterations, the number repair locations, and the repair time (in seconds). }\label{tab:exp-iter}
	\def\arraystretch{1.1}
	\setlength\tabcolsep{1.2mm}
	\centering
	\scalebox{0.9}{
		\begin{tabular}{l@{\hspace{4mm}}c@{\hspace{4mm}}c@{\hspace{4mm}}c}
			\toprule
			\textbf{Instance} & \textbf{$\#$Iter} & \textbf{$\#$Locations} & $\boldsymbol{t}$\\
			\midrule
			\textsc{edas}		&2 &1 & 2.5\\
			\arrayrulecolor{black!20}
			\midrule
			\textsc{csrf}		&2 &1 & 17.9 \\
			\midrule
			\textsc{log}	&1 &1 & 0.9 \\
			\textsc{log$'$}	&1 &1 & 1.0 \\
			\textsc{log$''$}	&1 &1 & 7.4 \\
			\midrule
			\textsc{atm}	&3 &2 & 4.2 \\
			\midrule
			\textsc{reviews}	&3 &2 & 18.5 \\
			\textsc{reviews$'$}	&3 &2 & 151.6 \\
			\arrayrulecolor{black}
			\bottomrule
		\end{tabular}
	}
\end{wraptable}

We first focus on \tool{}'s ability to find, often non-trivial, repair solutions using its iterative repair approach.
\Cref{tab:exp-iter} depicts an overview of the 5 benchmark families we consider (explained in the following).
For some of the benchmarks, we also consider small variants by adding additional complexity to the program.

\paragraph{EDAS.}

As already discussed in \Cref{sec:overview}, \tool{} is able to repair (a simplified integer-based version of) the EDAS example in \Cref{fig:edas} and derive the repairs in \Cref{fig:repair-o}.

\begin{figure}[!t]
\begin{minipage}[c]{.55 \textwidth}
\begin{exampleCode}
login(int password, bool attack) {
@@if (password == 366) {
@@@@if (attack == true) {
@@@@@@request = 2 // hidden request (unsafe)
@@@@} else {
@@@@@@request = 1 // user request (safe)
@@@@}
@@} else {
@@@@request = 0 // empty request
@@}
@@request = request (*\label{line:csrf1}*)
@@observe
}\end{exampleCode}
\end{minipage}
	\hfill
	\begin{minipage}[c]{.33 \textwidth}
		\begin{subfigure}{ \textwidth}
			\begin{codeBox}
				\vspace{-1mm}
\begin{exampleCode}[numbers=none,xleftmargin=0.4em]
request = 0
\end{exampleCode}
\vspace{-1mm}
			\end{codeBox}
			\subcaption{}
			\vspace*{4mm}
			\label{fig:csrfrepairA}
		\end{subfigure}
		\begin{subfigure}{ \textwidth}
			\begin{codeBox}	
				\vspace{-1mm}
\begin{exampleCode}[numbers=none,xleftmargin=0.4em]
if (password == 366) {
@@request = 1
} else {
@@request = request
}
\end{exampleCode}
\vspace{-1mm}
			\end{codeBox}
			\subcaption{}
			\vspace*{4mm}
			\label{fig:csrfrepairB}
		\end{subfigure}
	\end{minipage}
	\caption{A CSRF attack and repair candidates by \tool.}
	\label{fig:csrf}
\end{figure}

 \paragraph{CSRF.} 
Cross Site Request Forgery (CSRF) \cite{KhanCBGP14} attacks target web session integrity.
As an abstract example, consider the simple login program as shown in \Cref{fig:csrf} (left), where we leave out intermediate instructions that are not necessary to understand the subsequent repair. 
If the user attempts to log in and enters the correct password, we either set \ex{request = 1} (modeling a login on the original page), or \ex{request = 2} (modeling an attack, i.e., a login request at some untrusted website).
We specify that the \ex{request} should only depend on the (correctness of the) \ex{password}.
When repairing line~\ref{line:csrf1}, \tool{} first discoverers the trivial repair that always overwrites \ex{request} with a fixed constant (\Cref{fig:csrfrepairA}). 
However, in the second improvement iteration, \tool{} finds a better repair (\Cref{fig:csrfrepairB}), where the 
\ex{request} is only overwritten after a successful login. 
The potential attack request (\ex{request = 2}) is thus deterministically overwritten.

\begin{figure}[t]
	\begin{minipage}[c]{0.65 \textwidth}
\begin{exampleCode}
log(string password, string username, 
@@string date) {
@@if(password == userPassword){
@@@@// password flows to credentials
@@@@credentials = username + password
@@} else {
@@@@credentials = username
@@}
@@// then flows to info
@@info = date + credentials 
@@// then flows to LOG
@@LOG = info (*\label{line:bank1}*)
@@observe
} \end{exampleCode}
\end{minipage}%
\begin{minipage}[c]{.35 \textwidth}
	\begin{subfigure}[b]{\linewidth}
		\begin{codeBox}
			\vspace{-1mm}
\begin{exampleCode}[numbers=none,xleftmargin=0.4em]
LOG = ""
\end{exampleCode}
			\vspace{-1mm}
		\end{codeBox}
		\subcaption{}
		\label{fig:bankrepairA}
		\vspace*{4mm}
	\end{subfigure}%
	
	\begin{subfigure}[b]{\linewidth}
		\begin{codeBox}
			\vspace{-1mm}
\begin{exampleCode}[numbers=none,xleftmargin=0.4em]
LOG = date + username
\end{exampleCode}
			\vspace{-1mm}
		\end{codeBox}
		\subcaption{}
		\label{fig:bankrepairB}
		\vspace*{4mm}
	\end{subfigure}
\end{minipage}
\caption{Privacy leakage by logging and repair candidates by \tool. }
\label{fig:bank}
\end{figure}

\begin{wrapfigure}{R}{0.45\linewidth}
	\vspace{-10mm}
\begin{exampleCode}
atm(int balance, int amount) {
@@if (balance < amount){
@@@@ErrorLog = "overdraft"
@@} else {
@@@@balance = balance - amount
@@@@TransactionLog = "success"
@@}
@@ErrorLog = ErrorLog(*\label{line:atm1}*)
@@TransactionLog = TransactionLog(*\label{line:atm2}*)
@@observe
}
\end{exampleCode}
	\vspace{-2mm}
	\captionof{figure}{An ATM that leaks the \exm{balance} to \exm{ErrorLog} and \exm{TransactionLog}.\vspace{-4mm}}
	\label{fig:atm}
\end{wrapfigure}

\paragraph{LOG.} 

We investigate privacy leaks induced by \emph{logging} of credentials.
We depict a simplified code snipped in \Cref{fig:bank}.
Crucially, in case of a successful login, the secret \ex{password} flows into the public \ex{LOG} (via \ex{credentials} and \ex{info}). 
We specify that the \ex{LOG} may only depend on public information (i.e., everything except the \ex{password}) and use \tool{} to overwrite the final value of \ex{LOG} (i.e., to repair line~\ref{line:bank1}). 
As shown in \Cref{fig:bankrepairA}, \tool{} first finds a trivial repair that does not log anything.
In the first improvement iteration, \tool{} automatically finds the more accurate repair in \Cref{fig:bankrepairB}.
That is, it automatically infers that \ex{LOG} can contain the date and username (as in the original program) but not the password.

\begin{figure}[t]
	\begin{minipage}[c]{.65 \textwidth}
\begin{exampleCode}
reviews(int reviewerAid, int reviewerBid, 
@@string reviewA, string reviewB) {
@@notification = "Your CAV24 reviews:"
@@if (reviewerAid <= reviewerBid){
@@@@order = 1
@@} else {
@@@@order = 2
@@}
@@order = order(*\label{line:review1}*)
@@if (order == 1) {
@@@@notification = notification + reviewA
@@@@notification = notification + reviewB
@@} else {
@@@@notification = notification + reviewB
@@@@notification = notification + reviewA
@@}
@@observe
}
\end{exampleCode}
	\end{minipage}%
	\begin{minipage}[c]{.35 \textwidth}
		\begin{subfigure}[b]{\linewidth}
			\begin{codeBox}
				\vspace*{-1mm}
\begin{exampleCode}[numbers=none,xleftmargin=0.4em]
order = 0
\end{exampleCode}
				\vspace*{-1mm}
				
			\end{codeBox}
			\subcaption{}
			\label{fig:review1}
			\vspace*{4mm}
		\end{subfigure}
		\begin{subfigure}[b]{\linewidth}
			\begin{codeBox}
				\vspace*{-1mm}				
\begin{exampleCode}[numbers=none,xleftmargin=0.4em]
if (reviewerAid < 2) {
@@order = order
} else {
@@order = 2
}
\end{exampleCode}
				\vspace*{-1mm}
			\end{codeBox}
			\subcaption{}
			\label{fig:review2}
			\vspace*{4mm}
		\end{subfigure}
	\end{minipage}
	
	\caption{A review system that leaks the reviewer ids via the review order and repair candidates by \tool. }
\label{fig:review}
\end{figure}

\paragraph{ATM.}
\label{sub:multiple}
Many cases require repairing \emph{multiple} lines of code simultaneously.
We use cases derived from open-source security 
benchmarks~\cite{HamannHMM0T18,GordonKPGNR15,SecuriBench} and mark multiple repair locations 
in the input programs.  
For example, consider the ATM program in \Cref{fig:atm}. Depending on whether the withdraw 
\ex{amount} is greater than \ex{balance} (secret), different messages will be logged (public).
To repair it, we need to repair both \ex{ErrorLog} and \ex{TransactionLog} under different conditions (i.e., do not update \ex{ErrorLog} in the if-clause and do not update \ex{TransactionLog} in the else-clause).
By indicating lines \ref{line:atm1} and \ref{line:atm2} as two repair locations, \tool{} is able to 
synthesize the correct multiline repair.

\paragraph{REVIEWS.}
We also investigate the review system depicted in \Cref{fig:review} (left).
Here the \ex{id} of each reviewer determines in which \ex{order} the reviews are displayed to the author. 
We assume that the PC chair always has the fixed ID $1$ (so if he/she submits a review, it will always be displayed first).
We want to avoid that the author can infer which review was potentially written by the PC chair. 
When asked to repair line \ref{line:review1}, \tool{} produces the repair patches displayed in \Cref{fig:review1,fig:review2}.
In particular, the last repair infers that if \ex{reviewerAid < 2} (i.e., reviewer \texttt{A} is the PC chair), we can leave the order; otherwise, we use some fixed constant.

\begin{table}[!t]
	
	\caption{In \Cref{tab:scale-sol-size}, we evaluate \tool{}'s scalability in the SyGuS solution size. The timeout (denoted ``'-'') is 120 seconds.
		In \Cref{tab:k-safety}, we repair a selection of $k$-safety instances from
		\cite{FarzanV19,ShemerGSV19,UnnoTK21,BeutnerF22}.
		In \Cref{tab:func}, we evaluate on a selection of functional repair instances from 
		\cite{MechtaevYR16,GouesDFW12}. 
		All times are given in seconds.
	}
	
	\begin{subtable}{0.28\linewidth}
		\subcaption{}\label{tab:scale-sol-size}
		\def\arraystretch{1.2}
		\setlength\tabcolsep{1.2mm}
		\centering
		\small
		\begin{tabular}{l@{\hspace{2mm}}c@{\hspace{2mm}}c@{\hspace{2mm}}c}
			\toprule
			$\boldsymbol{n}$ & \textbf{\#Iter} & $\boldsymbol{t}$ & \textbf{Size} \\
			\midrule
			0 & 0  & 0.8 & 1 \\
			1  & 0 & 0.8 & 1 \\
			2 & 1  & 1.1 & 3\\
			3 & 2  & 1.4 & 5\\
			4 & 3  & 1.8 & 7\\
			5 & 4  & 5.1 & 9\\
			6  & 5 & 89.8 & 11\\
			7 & - & - & -\\
			\bottomrule
		\end{tabular}
		
	\end{subtable}%
	\begin{subtable}{0.4\linewidth}
		\subcaption{}\label{tab:k-safety}
		\def\arraystretch{1.2}
		\setlength\tabcolsep{1.2mm}
		\centering
		\small
			\begin{tabular}{l@{\hspace{3mm}}c}
				\toprule
				\textbf{Instance} & $\boldsymbol{t}$ \\
				\midrule
				\textsc{CollItemSym}   & 1.4 \\
				\textsc{CounterDet}   & 4.9 \\
				\textsc{DoubleSquareNiFF} & 4.2 \\
				\textsc{DoubleSquareNi}   & 2.9\\
				\textsc{Exp1x3}  & 1.1 \\
				\textsc{Fig2} & 2.4 \\
				\textsc{Fig3}   & 1.1 \\
				\textsc{MultEquiv} & 2.0 \\
				\bottomrule
			\end{tabular}
		
	\end{subtable}%
	\begin{subtable}{0.32\linewidth}
		\subcaption{}\label{tab:func}
		\def\arraystretch{1.2}
		\setlength\tabcolsep{1.2mm}
		\centering
		\small
			\begin{tabular}{l@{\hspace{3mm}}c}
				\toprule
				\textbf{Instance} & $\boldsymbol{t}$ \\
				\midrule
				\textsc{Assignment}   & 0.7 \\
				\textsc{Deletion}  & 0.7 \\
				\textsc{Guard}   & 0.6 \\
				\textsc{Long-Output}   & 0.7 \\
				\textsc{Multiline}   & 0.8 \\
				\textsc{Not-Equal}   & 0.6 \\
				\textsc{SimpleExample}   & 1.0 \\
				\textsc{OffByOne} & 2.1 \\
				\bottomrule
			\end{tabular}
		
	\end{subtable}

\end{table}

\subsection{Scalability in Solution Size}

Most modern SyGuS solvers rely on a (heavily optimized) \emph{enumeration} of solution candidates 
\cite{ReynoldsBNBT19,HuangQSW20,AlurRU17,DingQ24}.
The synthesis time, therefore, naturally scales in the size of the smallest solutions.
Our above experiments empirically show that most repairs can be achieved by small patches.
Nevertheless, to test the scalability in the solution size, we have designed a benchmark family that only 
admits large solutions. 
Concretely, we consider a program that computes the conjunction of $n$ Boolean inputs $i_1, \ldots, 
i_n$.
We repair against a simple $\forall^2$ HyperLTL property which states that the output may not depend on the last input, guiding the repair towards the optimal solution $i_1 \land \cdots \land
i_{n-1}$.
We display the number of improvement iterations, the run time, and the solution size (measured in terms of AST nodes) in 
\Cref{tab:scale-sol-size}. 
We note that one of the main features of SyGuS is the flexibility in the input grammar.
When using a 
less permissive (domain-specific) grammar, \tool{} scales to even larger repair solutions.

\subsection{Evaluation on $k$-Safety Instances}

To demonstrate that \tool{} can tackle the repair problem in the size-range supported by current 
\emph{verification} approaches for hyperproperties, we collected a small set of $k$-safety 
verification instances from 
\cite{FarzanV19,ShemerGSV19,UnnoTK21,BeutnerF22}.
We modify each program such that the $k$-safety property is violated and use \tool{}'s plain (non-iterative) SyGuS constraints to find a repair.
The results in \Cref{tab:k-safety} demonstrate that
(1) existing off-the-shelf SyGuS solver can repair programs of the complexity studied in the context 
of $k$-safety \emph{verification}, and (2) even in the presence of loops (which are included in all 
instances in \Cref{tab:k-safety}), finite unrolling often suffices to generate repair constraints that yield repair patches that work for the full program. 

\subsection{Evaluation on Functional Properties}

While we cannot handle the large programs supported by existing APR approaches for 
functional properties, we can evaluate \tool{} on (very) small test cases.
We sample instances from \texttt{Angelix} \cite{MechtaevYR16} and \texttt{GenProg} \cite{GouesDFW12},
and apply \tool{}'s
direct (non-iterative) repair. 
We report the run times in \Cref{tab:func}.

\section{Related Work}
\label{sec:related-work}

\paragraph{APR.}

Existing APR approaches for functional properties can be grouped into  search-based and 
constraint-based \cite{GouesPR19,GazzolaMM19}.
Approaches in the former category use a heuristic to explore a set of possible patch candidates.
Examples include \texttt{GenProg} \cite{GouesDFW12} and \texttt{PAR} \cite{KimNSK13}, 
\texttt{SPR} \cite{LongR16}, \texttt{TBar} \cite{LiuK0B19}, or machine-learning-based approaches 
\cite{ZhuSXZY0Z21,FanGMRT23}.
These approaches typically scale to large code bases, but might fail to find a solution (due to the 
large solution space).
Our approach falls within the latter (constraint-based) category.
This approach was pioneered by \texttt{SemFix} \cite{NguyenQRC13} and later refined by 
\texttt{DirectFix} \cite{MechtaevYR15}, \texttt{Angelix} \cite{MechtaevYR16}, and \texttt{S3} 
\cite{LeCLGV17}.
To the best of our knowledge, we are the first to employ the (more general) SyGuS framework for APR, which leaves the exact search to an external solver.
Most APR approaches rely on a finite set of input-output examples.
To avoid overfitting \cite{SmithBGB15} these approaches either use heuristics (to, e.g., infer variables that a repair should depend on  \cite{XiongWYZH0017}) or employ richer (e.g., MaxSMT-based) constraints \cite{MechtaevYR16}.
Crucially, these approaches are \emph{local}, whereas our repair constraints reason about the entire (\emph{global}) program execution by utilizing the entire symbolic path. 
Any repair sequence generated by our iterative repair is thus guaranteed to increase in quality, i.e., preserve more behavior of the original program.

\paragraph{APR for Hyperproperties.}

Coenen et al.~\cite{CoenenFHHS21} study \emph{enforcement} of alternation-free hyperproperties.
Different from our approach, enforcement does not provide guarantees on the functional behavior of the enforced system.
Bonakdarpour and Finkbeiner \cite{BonakdarpourF19} study the repair-complexity of hyperproperties in \emph{finite}-state systems.
In their setting, a repair consists of a substructure, i.e., a system obtained by removing some of the transitions of the system, so the repair problem is trivially decidable.
Polikarpova et al.~\cite{PolikarpovaSYIH20} present \texttt{Lifty}, and encoding of information-flow properties using refinement types.
\texttt{Lifty} can automatically patch a program to satisfy an information-flow requirement by 
assigning \emph{all} private variables some public dummy default constant. 
In contrast, our approach can repair against complex temporal hyperproperties (possibly involving quantifier alternations), and our repairs often go beyond insertion of constants.

\section{Conclusion}
\label{sec:concl}

We have studied the problem of automatically repairing an (infinite-state) software program against a 
temporal hyperproperty, using SyGuS-based constraint generation.
To enhance our basic SyGuS-based approach, we have introduced an iterative repair approach inspired by 
the notion of transparency.
Our approach interprets ``closeness'' rigorously, encodes it within our constraint system for 
APR, and can consequently derive non-trivial repair patches. 

\subsubsection*{Acknowledgments.}

This work was partially supported by the European Research Council (ERC) Grant HYPER (101055412), by the German Research Foundation (DFG) as part of TRR 248 (389792660), and by the United States NSF SaTC Awards 210098 and 2245114.

%
%
 \bibliographystyle{splncs04}
 \bibliography{references}
 
\end{document}